\documentclass[aoas]{imsart}

\RequirePackage{amsthm,amsmath,amsfonts,amssymb}
\RequirePackage[authoryear]{natbib}
\RequirePackage[colorlinks,citecolor=blue,urlcolor=blue]{hyperref}
\RequirePackage{graphicx}
\RequirePackage{mathtools,mathrsfs,url}
\newcommand{\PreserveBackslash}[1]{\let\temp=\\#1\let\\=\temp}

\newtheorem{theorem}{Theorem}
\newtheorem{lemma}{Lemma}
\newtheorem{example}{Example}

\newtheorem{algorithm}{Algorithm}

\newcommand{\reva}[1]{{\color{red} #1}}
\newcommand{\revb}[1]{{\color{blue} #1}}

\newcommand{\revd}[1]{{\color{red} #1}}

\renewcommand{\reva}[1]{{#1}}
\renewcommand{\revb}[1]{{#1}}
\renewcommand{\revd}[1]{{#1}}
\def\Unif{\mbox{Unif}}
\def\MSE {\mbox{MSE}}

\startlocaldefs

\endlocaldefs

\begin{document}

\begin{frontmatter}
\title{Orthogonal Subsampling for Big Data Linear Regression}
\runtitle{Orthogonal Subsampling}

\begin{aug}
\author[A]{\fnms{Lin} \snm{Wang}\ead[label=e1]{linwang@gwu.edu}},
\author[B]{\fnms{Jake} \snm{Elmstedt}\ead[label=e2,mark]{jake.r.kramer@gmail.com}}
\author[C]{\fnms{Weng Kee} \snm{Wong}\ead[label=e3]{wkwong@ucla.edu}}
\and
\author[B]{\fnms{Hongquan} \snm{Xu}\ead[label=e4,mark]{hqxu@stat.ucla.edu}}
\address[A]{Department of Statistics, George Washington University, \printead{e1}}

\address[B]{Department of Statistics, University of California, Los Angeles, \printead{e2,e4}}

\address[C]{Department of Biostatistics, University of California, Los Angeles, \printead{e3}}
\end{aug}

\begin{abstract}
The dramatic growth of big datasets presents a new challenge to data storage and analysis. Data reduction, or subsampling, that extracts useful information from datasets is a crucial step in big data analysis. We propose an orthogonal subsampling (OSS) approach for big data with a focus on linear regression models. The approach is inspired by the fact that an orthogonal array of two levels provides the best experimental design for linear regression models in the sense that it minimizes the average variance of the estimated parameters and provides the best predictions. The merits of OSS are three-fold: (i) it is easy to implement and fast; (ii) it is suitable for distributed parallel computing and ensures the subsamples selected in different batches have no common data points; and (iii) it outperforms existing methods in minimizing the mean squared errors of the estimated parameters and maximizing the efficiencies of the selected subsamples. Theoretical results and extensive numerical results show that the OSS approach is superior to existing subsampling approaches. It is also more robust to the presence of interactions among covariates and, when they do exist, OSS provides more precise estimates of the interaction effects than existing methods. The advantages of OSS are also illustrated through analysis of real data.
\end{abstract}


\end{frontmatter}

\section{Introduction}
Modern research in various areas is characterized by the unprecedented demand of big data analysis. As the scale of datasets increases, so does the demand of computational resources in the analysis and modeling process. Although the computational power has been developing rapidly, it still falls far behind under the explosive increase in data volume. Data reduction, or subsampling, that draws informative and efficient subsamples from big datasets for downstream modeling provides a feasible and principal way of converting big data into knowledge with available computational resources.

In many scientific research areas, the goal of data analysis is to study the effect of covariates on one or more response variables. Linear regression is  a primary tool for analyzing big data because there are many well developed tools for a linear model and the parameters are  straightforward to interpret.  For example,
\cite{melie2018multiple} proposed a linear regression methodology for the analysis of big data about human brains in the framework of the medical informatics platform;
\cite{saber2017short} presented short term load forecasting using linear regression for big data in power system, {and}
\cite{majumdar2017analysis} focused on linear regression for the analysis of big agriculture data with the goal of finding optimal parameters to maximize the crop production. Many feature selection methods are also linear in nature \citep{tibshirani1996regression, zou2005regularization}.
Even in applications where complex models are plausible, linear regression is {often still} employed as the first step to explore linear associations between covariates and one or more response variables.

For linear regression with an $n\times 1$ response vector and $n\times p$ covariate matrix, the required computing time for obtaining the ordinary least squares (OLS) estimator is $O(np^2+p^3)$. This time complexity is too long for big data, and may even be beyond the computational capacity of available computing facilities.
Therefore, despite the simplicity, linear regression in practice can be slow due to the large data volumes.
In addition, the limitation of storage capacity of individual's computing resources often makes it prohibitive to train the full dataset when it is too large. Training an efficient subsample instead of the full data can significantly accelerate the modeling process and tackle the issue of storage capacity.
Subsampling is then necessary in many applications and efficient subsampling approaches are thus in high demand.

Subsampling can be carried out to reduce either the dimensionality of datasets (i.e., the number of covariates $p$) or the number of data points (i.e., $n$).
When $p$ is large relative to $n$, the objective is to reduce the number of covariates while keeping $n$ fixed. This subsampling process is referred as dimension reduction or feature selection for high-dimensional data \citep{tibshirani1996regression, candes2007dantzig}. Recently, \cite{fan2020subsampling} developed a Subsampling Winner Algorithm that randomly subsamples the full data and selects features based on the scores of each feature obtained from subsample analyses.
In another scenario with $n\gg p$ and $n$ extremely large, the goal is to build a model that studies the effects of all covariates on the response variable. In this case, we take a subsample with $k$ data points without eliminating any covariates. Here $k$ can be predetermined based on computational capacity and is typically much smaller than $n$. Interesting literature in this direction includes \cite{drineas2012fast}, \cite{ma2015leveraging}, and \cite{wang2018information}, among others.

This paper focuses on the scenario with $n\gg p$, which is an important and common problem that arises in both scientific work and real life problems. For example, \cite{wang2018information} considered an airline on-time dataset from the 2009 ASA Data Expo which contains $n=123,534,969$ observations on $p=29$ variables about flight arrival and departure information for all commercial flights within the USA, from October 1987 to
April 2008. Two other problems are discussed in Section \ref{sec:realdata}, where we consider a wave energy converters dataset with $n=288,000$ observations and $p=48$ variables and a blog feedback dataset with $n=60,021$ observations and $p=280$ variables.
Computing resources to analyze the full datasets in these examples are challenging, and thus subsampling which extracts $k$ informative data points is a crucial step for the analysis. 
For the case when $n\gg p$, the term involving $n$ dominates the computing complexity, which implies that the required
computing time for the OLS estimator on the full sample is $O(np^2)$.
\cite{drineas2012fast} developed a randomized subsampling algorithm based on statistical leverage scores and reduced the computing time to at least $O(np\log n/\epsilon^2)$, where $\epsilon\in (0,0.5]$ is a predefined small constant. 
\cite{wang2018information} proposed an information-based optimal subdata selection (IBOSS) method and reduced the computing time to $O(np)$. They also showed, both theoretically and numerically, that the IBOSS method typically outperforms the leveraging scores method in selecting informative subsamples from big data.

In this paper, we propose a new subsampling approach, which we call orthogonal subsampling (OSS), for big data.
The approach is inspired by the fact that an orthogonal array of two levels provides the best experimental design for
linear regression in the sense that it minimizes the average variance of parameter estimations and provides the best predictions {at the same time} \revd{\citep{dey1999fractional}}.
Classical experimental design allows data points to be freely designed thus may not be a well suited tool for data-driven problems. Many design strategies, such as orthogonal arrays, however, can be borrowed to establish efficient approaches to select informative subsamples from big data.
The optimality of orthogonal arrays comes from their combinatorial orthogonality. The OSS approach, in an analogous manner, seeks subsamples with maximum combinatorial orthogonality.
To this end, we develop an algorithm that sequentially selects data points from the full sample with the aim of maximizing the subsample's orthogonality.
Using theory and numerical results, we show that there are at least three merits of the OSS approach. First, for a large $n$ and a fixed subsample size $k$, the computational complexity is $O(np\log k)$, which is as low as $O(np)$ from the IBOSS approach.
Second, the OSS is very suitable for distributed parallel computing and ensures that the subsamples selected in different batches have no common points.  Third, the OSS outperforms existing methods in minimizing the mean squared errors of the estimates for the model parameters and maximizing the efficiencies of the selected subsample. It is also robust to the presence of interactions and provides more precise estimates of the interaction effects than available methods.

The OSS approach is also suitable for measurement-constrained regression problems, where we are given a large pool of $n$ data points and can only observe a small set of $k$ responses (labels) due to the limit of time and budget.
These problems arise in a variety of areas in practice.  For example,
\cite{wang2017computationally} considered three such problems which concern material synthesis, CPU benchmarking, and wind speed prediction. In all the three applications, collecting responses for data points is either time-consuming or costly, so only a subset of data points can be observed. This is akin to an experimental design problem, but unfortunately, existing techniques of experimental design are either not applicable to a discrete subsample space or
suffer high computational cost, as discussed in Section 2.
Since the OSS approach only relies on the covariate matrix and applies to the discrete subsample space, it can be used naturally in the subsampling task for measurement-constrained problems and keep all the aforementioned merits.

In Section 2, we present the fundamental subsampling framework for the OSS method. In Section 3, we illustrate the optimality of orthogonal arrays. In Section 4, we propose the OSS method and use it to develop a computationally efficient algorithm for sequentially subsampling from big data.
In Section 5, we examine the performance of the OSS method through extensive simulations, discuss models with interactions, and demonstrate the utility of using the OSS method to \revd{select} subsamples for \revb{two} real big data sets.
We offer concluding remarks in Section 6 and show the proof of technical results in the supplementary materials \citep{Supplement2021}.

\section{The Framework}
Let $(x_1,y_1),\ldots,(x_n,y_n)$ denote the full sample, where $x_i=(x_{i1},\ldots,x_{ip})$ consists of the values of $p$ covariates and $y_i$ is the corresponding response. Assume the data follow the linear regression model:
\begin{equation}\label{slm}
y_i=\beta_0+\sum_{j=1}^{p}\beta_j x_{ij} + \varepsilon_i,
\end{equation}
where $\beta_0$ is the intercept, $\beta_1,\ldots,\beta_p$ are the slope parameters, and $\varepsilon_i$ is the independent random error with mean 0 and variance $\sigma^2$.
Let $y=(y_1,\ldots,y_n)^T$, $X=(x_{ij})$, $\tilde{X}=(1,X)$, $\beta=(\beta_0,\beta_1,\ldots,\beta_p)^T$, and $\varepsilon=(\varepsilon_1,\ldots,\varepsilon_n)^T$.
The matrix form of the model in \eqref{slm} is given by
\begin{equation}\label{c4lm}
y = \tilde{X}\beta + \varepsilon
\end{equation}
where the OLS estimator of $\beta$ is
$$
\hat{\beta} = (\tilde{X}^T\tilde{X})^{-1}(\tilde{X}^Ty).
$$
Now consider taking a subsample of size $k$ from the full sample $(X,y)$. Denote the subsample by $(X_s, y_s)$ which contains data points $(x_{s_1},y_{s_1}),\ldots,(x_{s_k},y_{s_k})$ from the full sample. Here $\{s_1,\ldots,s_k\}$ is a subset of $\{1,\ldots,n\}$. The OLS estimator based on the subsample is given by
$$
\hat{\beta}_s = (\tilde{X}_s^T\tilde{X}_s)^{-1}(\tilde{X}_s^Ty_s),
$$
where $\tilde{X}_s=(1,X_s)$. The $\hat{\beta}_s$ is the best linear unbiased estimator of $\beta$ based on the subsample $(X_s, y_s)$, that is, $\hat{\beta}_s$ has the minimum variance among all linear unbiased estimators based on the subsample $(X_s, y_s)$. The covariance matrix
of $\hat{\beta}_s$ is
\begin{equation}\label{ms}
\sigma^2M_s^{-1} = \sigma^2(\tilde{X}_s^T\tilde{X}_s)^{-1}.
\end{equation}
The covariance matrix in \eqref{ms} is a function of $X_s$, which indicates that the subsampling strategy is critical in reducing the variance of $\hat{\beta}_s$.
To minimize the variance of $\hat{\beta}_s$, we seek the subsample $X_s$ which, in some sense, minimizes $M_s^{-1}=(\tilde{X}_s^T\tilde{X}_s)^{-1}$. This is typically done, in experimental design strategy, by minimizing an optimality function of the matrix $M_s^{-1}$ \citep{kiefer1959optimum1,atkinson2007optimum}. Denote $\psi$ as the optimality function, then the idea is to solve the optimization problem:
\begin{equation}\label{optim}
X_s^*=\arg\min_{X_s\subseteq X}\psi(M_s^{-1}) = \arg\min_{X_s\subseteq X}\psi((\tilde{X}_s^T\tilde{X}_s)^{-1}).
\end{equation}

Choices for $\psi$ include the determinant and the trace, which are akin to the criteria of $D$-optimality and $A$-optimality for the selection of optimal experimental designs. The $D$-optimality is to minimize $\det(M_s^{-1})$, the determinant of $M_s^{-1}$, so that the volume of the confidence ellipsoid formed by $\hat{\beta}_s$ is minimized. The $A$-optimality is to minimize the trace of $M_s^{-1}$, so that the average variance of $\hat{\beta}_s$ is minimized.
To be rigorous, $\psi(M_s^{-1})$ is defined to be infinity when $M_s$ is singular.

The optimization problem in \eqref{optim} is not easy to solve. Exhaustive search for solving the problem in \eqref{optim} requires $O(n^kk^2p)$ operations, which is infeasible for even moderate sizes for $X$ and $X_s$.
\revd{There are many types of algorithms for finding optimal designs and among them, exchange algorithms are among the most popular.  An early review of such algorithms is \cite{nguyen1992review} and a detailed description with computer codes of the Fedorov exchange algorithm for finding a $D$-optimal design is given in \cite{miller1994algorithm}.}
To apply the exchange algorithm, one starts from a random subsample, $X_s$, of size $k$ and exchange a point in $X_s$ with a point in the full sample so that a delta function that measures the increase in $\det(M_s)$ is maximized. The exchange continues until the increase in  $\det(M_s)$ is less than a chosen positive small number.  The number of operations involved in each exchange is $nkp^2+p^3$, which is even more than the operation required in linear regression on the full data, not to mention that the number of exchanges may also increase dramatically with $n$. Consequently, exchange algorithms are cumbersome in solving the subsampling problem in \eqref{optim}.
There are improvements in exchange algorithms, such as the coordinate exchange \revd{\citep{overstall2017bayesian} and column exchange \citep{li1997columnwise}} algorithms; 
\revd{however, due to the computational burden of exchange algorithms, they can perform very slowly or not converge at all, see for example \cite{chaloner1989optimal},  \cite{royle2002exchange} and \cite{xu2002algorithm}.}


A competent subsampling method should contain a feasible solution to the optimization problem in \eqref{optim}.
\cite{wang2017computationally} considers a continuous relaxation of the problem in \eqref{optim}, focusing on $A$-optimality, to obtain a nearly `optimal' probability distribution on which a stochastic subsample is then drawn from the full sample;
\cite{wang2018information} proposes an information-based optimal subdata selection (IBOSS) method that approximates the $D$-optimality by only selecting data points with extreme (largest and smallest) covariate values into the subsample.
Both methods try to select informative individual data points, but neither approximates the combinatorial optimality in \eqref{optim} because they do not deal with the `mutual restraint' among data points.
In fact, data points in a big dataset usually possess overlap information, so if a data point is included in the subsample, its `similar' point should not be included so that the subsample covers diverse information, by `similar' here we mean data points sharing much overlap information. This `mutual restraint' among data points contributes the combinatorial nature of the problem in \eqref{optim} and requires much computation to deal with.
The OSS approach we are presenting in this paper targets at selecting `dissimilar' data points to approximate the combinatorial optimization. Instead of direct search for the optimization in \eqref{optim}, we demonstrate the exact optimality of orthogonal arrays and therefore propose to select the subsample that best approximates an orthogonal array.


\section{Orthogonal Array}
An orthogonal array of strength 2 on $q$ levels is a matrix with combinatorial orthogonality, that is, entries of the matrix come from a fixed finite set of $q$ levels (symbols), arranged in such a way that all ordered pairs of the levels appear equally often in every selection of two columns of the matrix. Readers are referred to \cite{hedayatorthogonal} for a comprehensive introduction of orthogonal arrays. In this paper, we set $q=2$ and denote the two levels by $-1$ and 1. There are thus four possible ordered pairs in any two columns, which are $(-1,-1), (-1,1), (1,-1),$ and $(1,1)$, and they appear equally often.
Here is an example of $4\times 3$ orthogonal array where each ordered pair appear once:
$$
\begin{pmatrix*}[r]
-1 & -1 & -1 \\
-1 & 1 & 1 \\
1 & -1 & 1 \\
1 & 1 & -1
\end{pmatrix*}.
$$

The combinatorial orthogonality of orthogonal arrays is in fact a type of balance which ensures that all columns are considered equally and rows distributed dissimilarly to cover diverse information.
\cite{cheng1980orthogonal} showed that an orthogonal array of strength 2 on $q$ levels is universally optimal, i.e., optimal under a wide variety of criteria ($\psi$) that include $D$- and $A$-optimality, among all $q$-level factorial designs. Specifically, a two-level orthogonal array of strength 2 is $D$- and $A$-optimal among all two-level factorial designs for the first-order linear model in \eqref{slm}; see also \cite{dey1999fractional}.
The optimality of orthogonal arrays can be \revb{carried over} to the subsampling problem. Because a subsample
has to be a subset of the full sample, each covariate is bounded by its minimum and maximum in the full sample. Hence, we can always scale each covariate to $[-1,1]$.
If a scaled subsample $X_s$ forms an orthogonal array on the two levels $\{-1,1\}$, its combinatorial orthogonality ensures that $M_s=\tilde{X}_s^T\tilde{X}_s=kI$, where $I$ is the identity matrix of order $p+1$, so that $X_s$ would be $D$- and $A$-optimal in the subsampling problem in \eqref{optim}. We summarize this result as follows.

\begin{lemma}\label{lem1}
Suppose all covariates are scaled to $[-1,1]$. If a subsample $X_s$ forms a two-level orthogonal array of strength $2$, $X_s$ is $D$- and $A$-optimal for the model in \eqref{slm}.
\end{lemma}

Lemma \ref{lem1} indicates that a subsample is $D$- and $A$-optimal if it, when scaled to $[-1,1]$, forms a two-level orthogonal array of strength 2.
Our objective of subsampling is to find a subset of the full sample that forms an orthogonal array. However,
two issues may arise with the orthogonal array subsample. First, the number of rows of a two-level orthogonal array of strength 2 has to be a multiple of 4, meaning that an orthogonal array subsample is possible only when the subsample size $k$ is a multiple of 4. Second, even if $k$ is a multiple of 4, a full sample generally does not contain a subset of $k$ points which forms an orthogonal array. Therefore, finding an exact orthogonal array subsample may be impossible in many cases.
That being said, the optimality of orthogonal arrays \reva{inspires us to find a subsample that best approximates an orthogonal array to achieve near optimality for the problem in \eqref{optim}.}

\section{The OSS Approach}
Suppose all covariates are scaled to $[-1,1]$. The optimality of orthogonal arrays comes from two features of their arrangement of row points: (i) Extreme values: points are located at the corners of the data domain ($[-1,1]^p$) and have large distances from the center; and (ii) Combinatorial orthogonality: points (or precisely, their signs) are as dissimilar as possible. Feature (i) is consistent with the principle in optimal experimental design for the first-order regression model in \eqref{slm}, where design points are typically located at the boundary of the design domain.
{When the underlying model is (nearly) correctly specified, these extreme data points provide more information about the model.
Note that the IBOSS approach searches subsample points with only feature (i) in some sense without any consideration of feature (ii). This may result in a non-optimal subsample with imbalance among covariates or sample points.}

Our OSS approach minimizes a discrepancy function which measures the distortion of data points on keeping the two features simultaneously. Denote $x_i^*$, $i=1,\ldots,k$, as the $i$th data point in a subsample.
The discrepancy function contains two parts targeting the two features.
For feature (i), the function can include $p-\|x_i^*\|^2$ where $\|\cdot\|$ denotes the Euclidean norm. The constant $p$ is to ensure that the term $p-\|x_i^*\|^2$ is positive.
For feature (ii),
define
\revb{
\begin{equation}\label{c4delta}
\delta(x_i^*,x_j^*)= \sum_{l=1}^{p}\delta_1(x_{il}^*,x_{jl}^*)
\end{equation}
where $\delta_1(x,y)$ is equal to 1 if both $x$ and $y$ have the same sign (i.e., $x y>0$) and 0 otherwise.}
Then $\delta(x_i^*,x_j^*)$ is the number of components in $x_i^*$ and $x_j^*$ that have the same signs. {Combining} the two parts, we define a discrepancy function by:
\begin{equation}\label{c4D}
L(X_s) = \sum_{1\leq i<j \leq k} \big[p-\|x_i^*\|^2/2-\|x_j^*\|^2/2+\delta(x_i^*,x_j^*)\big]^2.
\end{equation}
We search for the sample $X_s$ that minimizes $L(X_s)$. That is, {our OSS methodology solves the optimization problem:}
\begin{eqnarray}\label{optim2}
  X_s^* &=& \arg\min_{X_s} L(X_s), \nonumber\\
   &s.t.& X_s \mbox{ contains } k \mbox{ points. }
\end{eqnarray}
The minimization of $L(X_s)$ is justified by the following metric property.
\begin{theorem}\label{c4jeq}
For a $k\times p$ subsample $X_s$,
$$
L(X_s) \geq \frac{k^2p(p+1)-4kp^2}{8},
$$
with equality if and only if $X_s$ forms a two-level orthogonal array.
\end{theorem}
Theorem \ref{c4jeq} shows that the discrepancy of $X_s$ has a lower bound which is attained if and only if $X_s$ forms an orthogonal array. In this sense, $L(X_s)$ can be viewed as a metric on the distortion of $X_s$ from an orthogonal array. The solution from \eqref{optim2} can then be interpreted as a subsample which best approximates an orthogonal array.

There are other ways to define a discrepancy function whose minimization would also result in an approximate orthogonal array. For example, because $M_s=\tilde{X}_s^T\tilde{X}_s=kI$ when $X_s$ is an orthogonal array, one can define the discrepancy function as \reva{the difference between the sum of squares of the off-diagonal entries in $M_s$ and the sum of the diagonal entries of $M_s$}.
In fact, this idea has contributed extensive literature in the field of experimental design for constructing the so-called supersaturated designs \revb{\citep{wu1993construction, cheng1997es2, xu2009recent}}. However, such a defined discrepancy may not be suitable for the subsampling task.
The exhaustive search for \revb{the minimization} requires $O(n^kk^2p)$ operations, which is infeasible for even moderate sizes for $X$ and $X_s$. The operations required are again more than the operations involved in the linear regression for the full sample. To the best of our knowledge, exchange algorithms are the only possible available algorithms for minimizing this discrepancy, but as discussed in Section 2, they are generally inefficient for subsampling.

A key merit of the discrepancy function in \eqref{c4D} is that it enables sequential minimization to speed up the search. Here we propose an algorithm to sequentially search for eligible data points and quickly minimize the discrepancy in \eqref{c4D}. 
Ineligible data points are eliminated from $X$ along the way to speed up the search.

We first discuss the selection of a new point given that some data points are already chosen to be included in a subsample.
Assume the algorithm is at the $i$th iteration where $X_s^{i}$ is the new matrix obtained by adding $x_i^*$ to $X_s^{i-1}$, $i=2,\ldots,k$. Then by \eqref{c4D},
$$
L(X_s^i) = l(x_i^*|X_s^{i-1})+L(X_s^{i-1})
$$
where 
\begin{equation}\label{relaloss}
l(x_i^*|X_s^{i-1})
=\sum_{x_j^*\in X_s^{i-1}} \big[p-\|x_i^*\|^2/2-\|x_{j}^*\|^2/2+\delta(x_i^*,x_{j}^*)\big]^2
\end{equation}
is the discrepancy introduced by adding $x_i^*$ to $X_s^{i-1}$. To minimize $L(X_s^i)$ when $X_s^{i-1}$ is already selected and fixed, we select $x_i^*$ which minimizes the discrepancy in \eqref{relaloss}, that is,
\begin{equation}\label{c4xstar}
x_i^* = \arg\min_{x\in X} ~l(x|X_s^{i-1}).
\end{equation}
Because $l(x|X_s^{i-1})=l(x|X_s^{i-2})+l(x| x_{i-1}^*)$ and we have calculated
$l(x|X_s^{i-2})$ when searching for $x_{i-1}^*$ at the $(i-1)$th iteration, we only need to compute
\begin{equation}\label{c4dscore}
l(x| x_{i-1}^*) = \big[p-\|x\|^2/2-\|x_{i-1}^*\|^2/2+\delta(x,x_{i-1}^*)\big]^2
\end{equation}
at the $i$th iteration.
The computational complexity of each iteration is $O(np)$. To reduce the computation, we delete some data points in $X$ with large values of $l(x|X_s^{i-1})$ so that these points will not be considered at the $(i+1)$th iteration.


The following algorithm outlines the steps for solving the optimization problem in \eqref{optim2}.
\begin{algorithm}[Sequentially selecting an OSS subsample]\label{alg}\rm\
\begin{itemize}
  \item[Step 1.] [Initiation] Scale values of each covariate to $[-1,1]$. Let $i=1$. Find the point  $x_1^*$ in $X$ with the largest Euclidean norm, include it in $X_s$ and remove it from $X$. {Let $\mathscr{L}$ be an $(n-1)$-vector with each component corresponding to each remaining data point in $X$. Set all components of $\mathscr{L}$ to be 0.}
  \item[Step 2.] [Election] Increase $i$ by 1. For each $x \in X$, add
$l(x| x_{i-1}^*)$ in \eqref{c4dscore}
to its corresponding component in $\mathscr{L}$. Find $x_i^*$ with the smallest component in $\mathscr{L}$ and add it to $X_s$.
  \item[Step 3.] [Elimination]
  For each iteration $i$, select a value for a parameter  $t_i$ that depends on $n$ and $k$. Then keep $t_i$  points in $X$ with $t_i$ smallest components in $\mathscr{L}$. Remove $x_i^*$ and other points from $X$ as well as their corresponding components from $\mathscr{L}$.
  \item[Step 4.] [End] Iterate Steps 2 and 3 until $X_s$ contains $k$  points.
\end{itemize}
\end{algorithm}

The parameter $t_i$ in the Elimination Step depends on the ratio between $n$ and $k$. If $n\gg k$, say $n>k^2$, then we have far more candidate points than needed so we can eliminate a large proportion of points and set $t_i$ to be much smaller than $n$; otherwise, we only eliminate a small portion and set $t_i$ to be close to $n$. For our numerical results in Section \ref{sec:numeric}, we set $t_i=n/i$ if $n\geq k^2$ and $t_i=n/i^{r-1}$ if $n<k^2$, where $r=\log(n)/\log(k)$.
As for the computational complexity, we are most interested when $n$ is large. In this case, $X$ consists of $t_i=n/i$ points at each iteration and the time for finding $x_i^*$ becomes $O(np/i)$. Therefore, the complexity for selecting $k$ data points is $O(np/1)+\cdots+O(np/k)= O(np\log k).$ Note that $k$ can be any integer less than $n$ in the OSS approach, not restricted to be a multiple of 4.

The form of the discrepancy function in \eqref{c4D} (and in \eqref{c4dscore}) may raise a question on the selection of center points with OSS. For example, let the $l$th components of three points, $x_i$, $x_j$ and $x_{j'}$, be  $x_{il}=0.01$, $x_{jl}=-0.01$, and $x_{j'l}=1$. Suppose $x_i$ has been selected into the subsample and we now need to make a choice between $x_j$ and $x_{j'}$. The $l$th component has a smaller contribution to $l(x_j|x_i)$ than to $l(x_{j'}|x_i)$ so favors the choice of $x_j$ rather than $x_{j'}$, which may not seem to make sense because $x_i$ and $x_{j}$ are closer. This concern seems to be plausible, but the OSS would typically not fall into that awkward situation. The key is on the understanding of the behaviour of the OSS algorithm. In each iteration,
the OSS selects the point with the largest norm in a quadrant that has not been covered (that is, none of the selected points is from this quadrant) and deletes points with small norms in the quadrant.
Typically, only the point with the largest norm in every quadrant may be included in the subsample considering the number of quadrants $2^p$. Therefore, if $x_i$ and $x_j$ are truly close to the center (and close to each other), they would be deleted early and would not even be considered as candidates for the subsample. This is why the function in \eqref{c4dscore} does not explicitly measure the discrepancy between center points and instead focuses on the points with big norms.

The OSS approach is well suited for distributed and parallel computing because the combination of orthogonal arrays is still an orthogonal array. One can separate the full dataset into $g$ {batches (groups)} and apply Algorithm \ref{alg} simultaneously on the batches, from each of which $k/g$ data points are selected.  The combination of all selected points would be the selected subsample from the full data. Such a subsample contains $k$ points that approximate an orthogonal array.

The following toy example illustrates the subsample selected by Algorithm \ref{alg} and its efficiency.
\begin{example}\label{c4toy}\rm
Suppose the full sample has $n=1000$ points,  $p=2$ covariates and we wish to select an informative subsample of size {$k=20$.}  The full sample is randomly generated from a bivariate uniform distribution on $[-1,1]^2$, that is, $x_i\sim \Unif[-1,1]^2$.
Figure \ref{toyexp} shows the full sample and subsamples selected by the UNI (uniform subsampling), IBOSS \citep{wang2018information}, WYS \citep{wang2017computationally}, and the OSS method. The UNI chooses points completely at random, IBOSS chooses boundary points, while the WYS and OSS methods choose data points near the corners. 
\begin{figure}
  \centering
  \includegraphics[width=.7\textwidth]{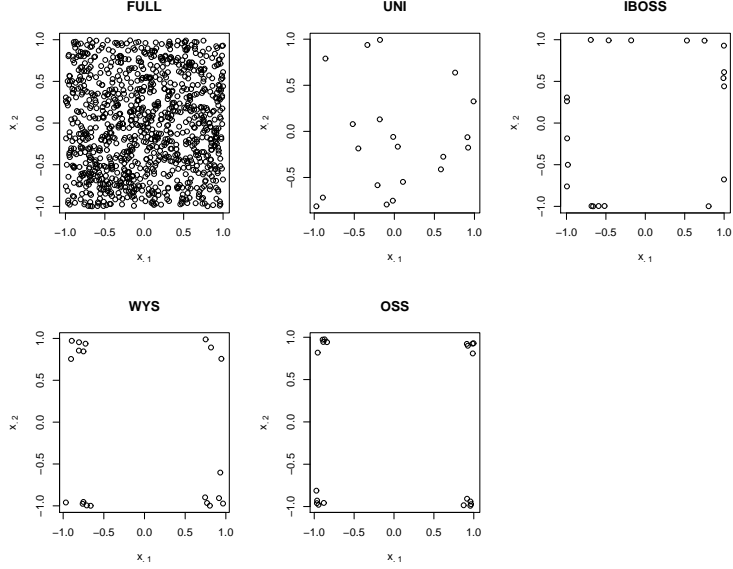}
  \vspace{-3mm}
  \caption{The full sample and subsamples selected by {different} methods.}\label{toyexp}
\end{figure}
To evaluate the performance of different subsamples, we randomly generated $T=100$ full samples (that is, $T$ repetitions) and calculated the $D$- and $A$-efficiencies of the subsamples selected by different methods. Because $\det(M_s)$ and $\sum_{j=0}^{p}\lambda_j(M_s^{-1})$ are minimized when $X_s$ is a two-level orthogonal arrays of strength 2, we have the following lower bounds:
$$
\det(M_s^{-1})\geq \frac{1}{k^{p+1}},\mbox{ and }\sum_{j=0}^{p}\lambda_j(M_s^{-1}) \geq \frac{p+1}{k}.
$$
The $D$- or $A$-efficiencies of a subsample can then be calculated using
\begin{equation}\label{c4deff}
D_{\rm eff}=\left\{(1/k^{p+1})/[\det(M_s)]^{-1}\right\}^{1/(p+1)}=\det(M_s)^{1/(p+1)}/k
\end{equation}
or
$$
\revb{A_{\rm eff}= [(p+1)/k]/\sum_{j=0}^{p}\lambda_j(M_s^{-1})
= (p+1)/\left[k\sum_{j=0}^{p}\lambda_j(M_s^{-1})\right] .}  
$$
In addition, for each subsample, we \revb{generated} the response variable $y$ through the model
$$
y_i = 1 + x_{i1} + x_{i2} + \varepsilon_i,
$$
\revb{where} $\varepsilon_i\sim N(0,9)$ for $i=1,\ldots,1000$.
The empirical mean squared {error} (MSE) is measured by
\begin{equation}\label{c4mse}
\MSE=T^{-1}\sum_{t=1}^{T}\|\hat{\beta}^{(t)}-\beta\|^2,
\end{equation}
where $\hat{\beta}^{(t)}$ is the OLS estimate of $\beta$ in the $t$th repetition based on the subsamples.
{Figure \ref{toymse} shows the MSE, $D$-efficiency, and $A$-efficiency for the subsamples selected by different approaches. The WYS and OSS methods outperform other approaches in each of the criteria because corner points are more informative for linear models and provide more accurate estimates for the model parameters. In addition, the OSS method selects corner points with combinatorial orthogonality and provides a better performance than the WYS method.}

\begin{figure}
  \centering
  \includegraphics[width=\textwidth, height=2.5in]{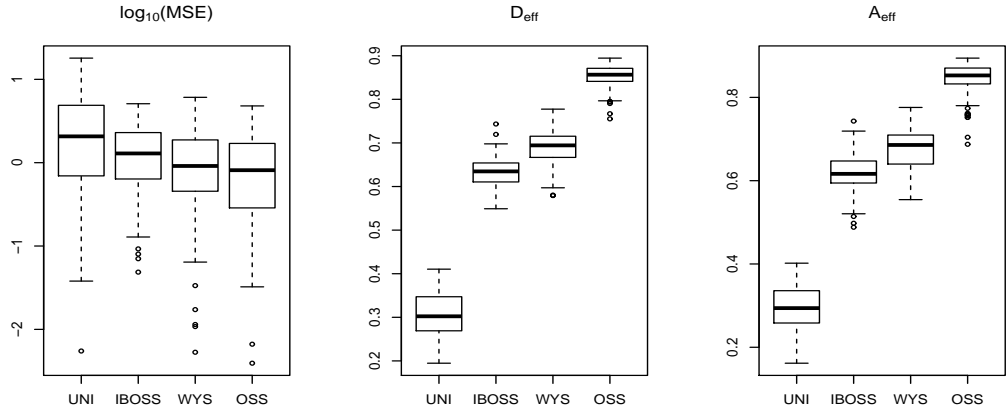}
  \vspace{-1cm}
  \caption{The MSE, $D$- and $A$-efficiencies for the subsamples selected by different methods.}\label{toymse}
\end{figure}
\end{example}

Example \ref{c4toy} is a toy example that suggests the OSS method outperforms the other three methods. Applications in the next section provide more numerical results that show the OSS method performs much better for larger $n$ and $p$.
Figure \ref{toyexp} may raise questions about potential outliers because it seems that the selected subsample by the OSS method only captures extreme covariate values. The IBOSS and WYS methods also have the same issue.
However, we show in Section \ref{sec:numeric} that this is not the case for a moderate or large $p$.  As $p$ increases, \reva{the selected subsample increasingly covers nonextreme points, so it allows the identification of outliers when diagnostic methods are used. If some of the selected points are outliers,
they are very likely to be identified in the subsample \revb{because} the selected subsample follows the same underlying regression model as the full data.}
On the other hand, if these corner points still follow the underlying model, then these data points actually
contain more information about the model and should be used for parameter estimation.

\section{Numerical Results}\label{sec:numeric}
In this section, we evaluate the performance of the OSS method using simulated and real data.

\subsection{Simulation Studies}
We conduct simulation studies to assess merits of the OSS method relative to {existing} subsampling schemes. Our setup is similar to that considered in \cite{wang2018information} for studying the IBOSS method. We use  model \eqref{c4lm}
and generate values of the covariates under three covariance structures:
\begin{itemize}
  \item[Case 1.] The covariates $x_i$'s are independent and have a multivariate uniform distribution with all covariates independent.
  \item[Case 2.] The covariates $x_i$'s have a multivariate normal distribution: $x_i\sim N(0,\Sigma)$, with
  \begin{equation}\label{Sigm}
  \Sigma=\left(0.5^{\xi(i,j)}\right),
  \end{equation}
  where \revb{$\xi(i,j)$ is equal to 1 if $i=j$ and 0 otherwise}.
  \item[Case 3.] The covariates $x_i$'s have a truncated multivariate normal distribution with mean 0 and variance matrix $\Sigma$ in \eqref{Sigm} and their values are constrained to $[-5,5]^p$.
\end{itemize}
The response data are generated from the linear model in \eqref{c4lm} with the true value of $\beta$ being a 51 dimensional vector of unity and $\sigma^2=9$. An intercept is included so $p=50$.

The simulation is repeated $T=1000$ times. We compare three subsampling approaches: UNI, IBOSS, and OSS. We do not include the method WYS for comparisons in the simulations because
it requires the computation of $X^T\mbox{diag}(\pi) X$ at each iteration, where $X$ is the full sample and $\pi$ is a vector of sampling probabilities. If the algorithm requires $n_{\rm iter}$ iterations, the complexity of such computation is $O(n_{\rm iter}(np^2+p^3))$,
which is roughly $n_{\rm iter}$ times the complexity of linear regressions on the full sample, so the WYS method loses its main utility if the goal is to accelerate the big data analysis. In addition, the Matlab package provided by the authors of the WYS method requires the storage of an $n\times n$ matrix, making the method unable to tackle the subsampling problem effectively when $n$ is big, say, $n\geq10^5$.

\begin{figure}
  \centering
  \includegraphics[width=\textwidth]{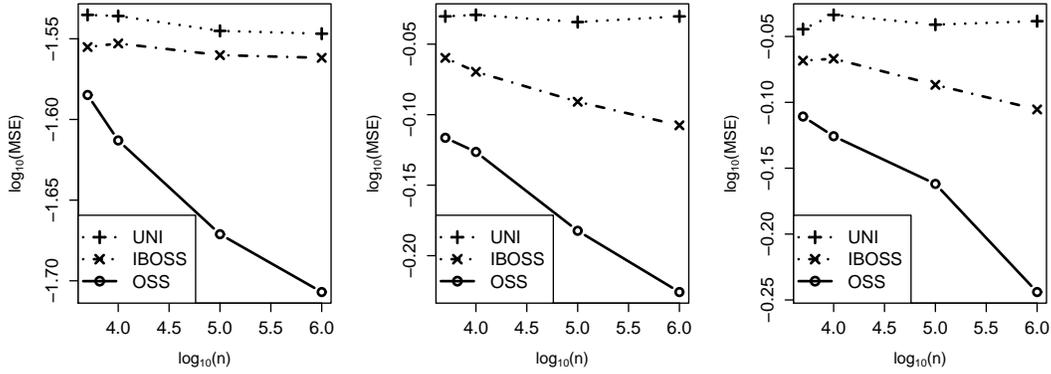}
  \caption{MSEs of the estimated slope parameters for three different distributions of the covariates: Case 1: uniform distribution (left), Case 2: normal distribution (middle), and Case 3: truncated normal distribution (right). The subsample size is fixed at $k=1000$.}\label{simumse}
\end{figure}

For each approach of UNI, IBOSS, and OSS, empirical MSEs are calculated for evaluation. As shown in \cite{wang2018information}, the IBOSS does not directly provide accurate estimation on the intercept parameter, the authors proposed using the adjusted intercept estimate:
\begin{equation}\label{intercept}
\hat{\beta}_0=\bar{y}-\bar{x}^T \hat{\beta}_{-0},
\end{equation}
where $\bar{y}$ is the mean of $y$, $\bar{x}$ is the vector of means of all variables in the full sample, and $\hat{\beta}_{-0}$ is the OLS estimate of $\beta_{-0}$ which includes all parameters except for the intercept based on the subsamples. To make a fair comparison, we estimate the intercept with \eqref{intercept} for all subsampling methods and separate its MSE from the MSE for the slope parameters. That is, we consider
$$
\MSE_{\beta_0}=T^{-1}\sum_{t=1}^{T}\|\hat{\beta}^{(t)}_0-\beta_0\|^2,
$$
and
$$
\MSE_{\beta_{-0}}=T^{-1}\sum_{t=1}^{T}\|\hat{\beta}^{(t)}_{-0}-\beta_{-0}\|^2,
$$
for the intercept and slope parameters separately, where $\hat{\beta}^{(t)}_{0}$ and $\hat{\beta}^{(t)}_{-0}$ are $\hat{\beta}_{0}$ and $\hat{\beta}_{-0}$ at the $t$th repetition.

We investigate several cases when the full sample sizes are $n=5\times10^3, 10^4, 10^5,$ and $10^6$ and the subsample size is fixed at $k=1000$. Figure \ref{simumse} presents plots of the $\log_{10}$ of the $\MSE_{\beta_{-0}}$ against $\log_{10}(n)$ for {model \eqref{c4lm}} with three covariance structures. We observe that the OSS method consistently outperforms the UNI and IBOSS methods for estimating the slope parameters. More importantly, the $\MSE_{-\beta_0}$ from the OSS method decreases as the full data sample size $n$ increases, even though the subsample size is fixed at $k=10^3$. This trend demonstrates that the OSS method extracts more information from the full sample when the full sample size increases. The $\MSE_{-\beta_0}$ from the IBOSS method is not decreasing in Cases 1 and 3 as $n$ increases because the covariates are bounded. This is consistent with the theoretical result in \cite{wang2018information} that $\MSE_{-\beta_0}$ from IBOSS decreases as $n$ increases only when covariates are unbounded.
We can see a decreasing trend of IBOSS in Case 2 with, however, a slower rate than OSS because the $\MSE_{-\beta_0}$ for IBOSS decreases fast mainly for unbounded covariates with heavy-tail distributions (say, the $t$ distribution with a low degree of freedom). Such distributed covariates may not be as common as bounded or normally distributed covariates in real applications, and when covariates do have heavy-tail distributions, it is common to consider suitable transformations of the covariates before a linear model is fitted. 
Figure 1 in the supplementary materials \citep{Supplement2021} shows that $\MSE_{\beta_0}$ does not change much across the three methods.

\begin{figure}
  \centering
  \includegraphics[width=.32\textwidth]{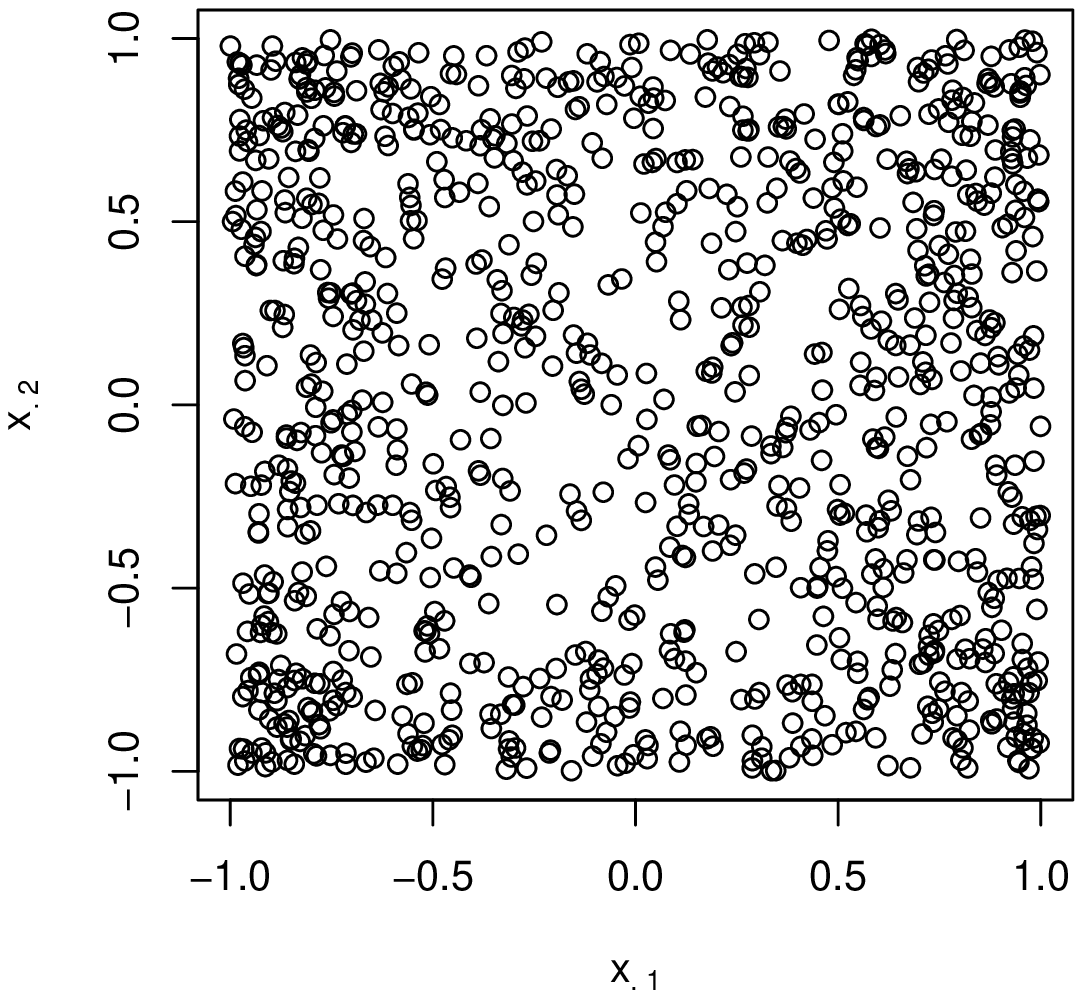}\includegraphics[width=.32\textwidth]{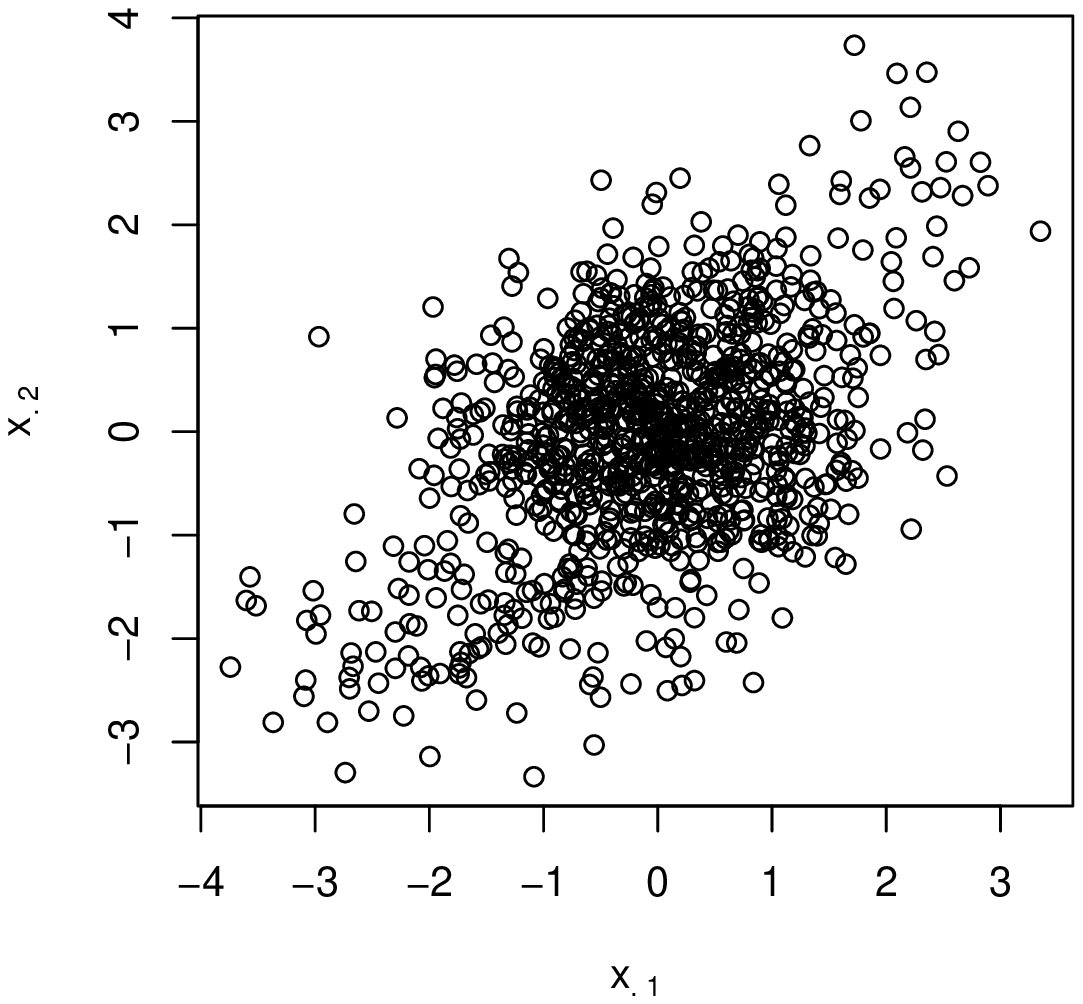}\includegraphics[width=.32\textwidth]{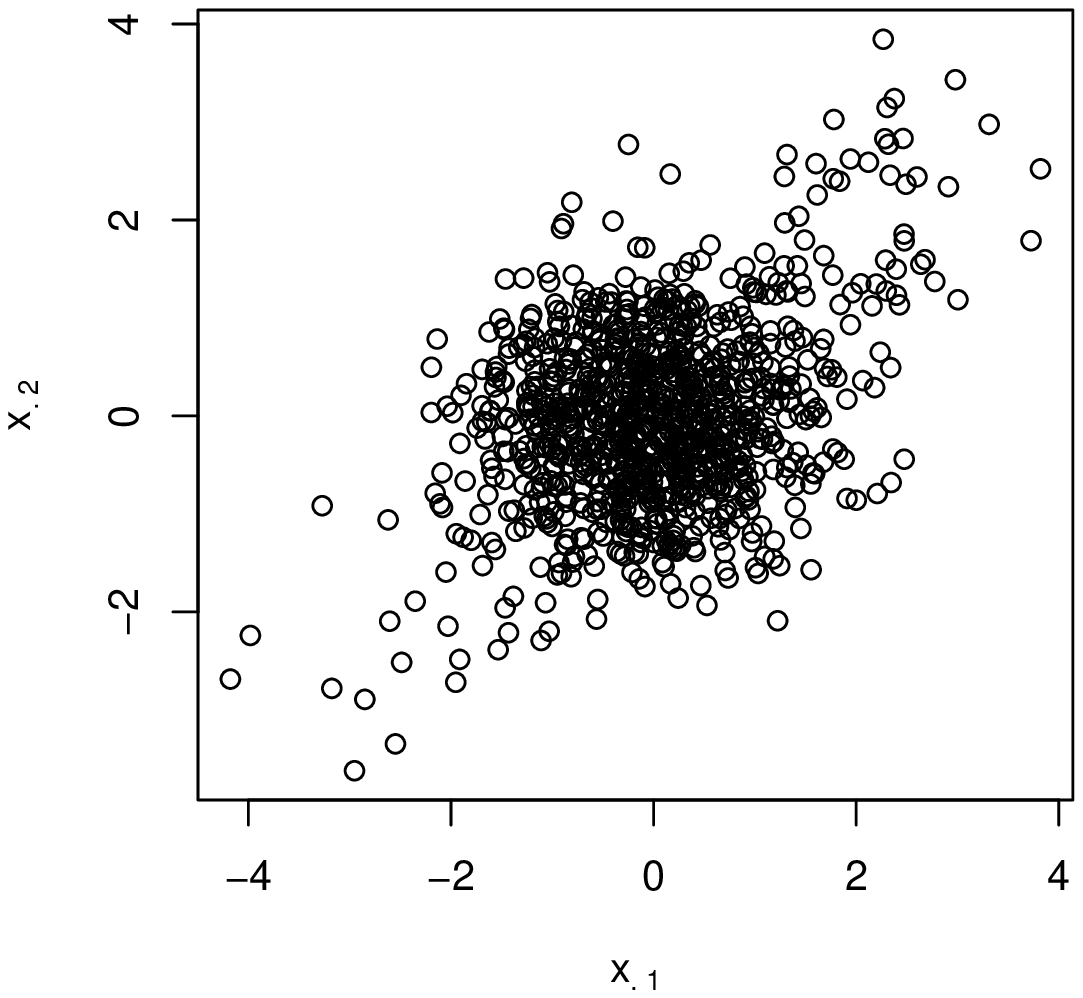}
  \vspace{-3mm}
  \caption{Two dimensional projection plots of the subsamples selected by the OSS approach for three different distributions for the covariates: Case 1: uniform distribution (left), Case 2: normal distribution (middle), and Case 3: truncated normal distribution (right). }\label{proj}
\end{figure}

Figure \ref{proj} shows the two dimensional projection plot of the subsamples selected by the OSS approach when $n=10^4$. {We only display the projection onto the first two covariates because projections onto other pairs of covariates would be similar.} We observe that the selected points are not concentrated at the corners, which is different from the case in Example \ref{c4toy} for $p=2$
and suggests that the selected subsample allows the identification of outliers when diagnostic methods are used.

\subsection{Models with Interactions}
\begin{figure}
  \centering
  \includegraphics[width=\textwidth]{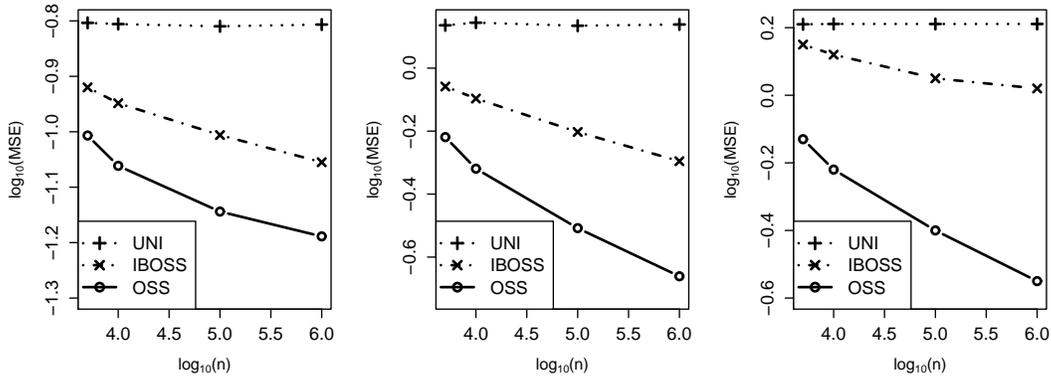}
  \caption{\revb{MSEs of the estimated parameters when the covariates follow a multivariate normal distribution (Case 2) and interactions exist: $\MSE_{\tilde{\beta}_1}$ (left) and $\MSE_{\tilde{\beta}_2}$ ({middle})  when the first-order plus interaction model is fitted, and  $\MSE_{\tilde{\beta}_1}$ (right) when the first-order model is fitted.} The subsample size is fixed at $k=1000$.}\label{interact}
\end{figure}

Interactions commonly exist among covariates.
To evaluate the performance of the OSS approach on the existence of interactions, we consider the situation where interactions exist between covariates. {For each covariate scenario} in the simulation, we generate the response data from the model
\begin{equation}\label{c4inter}
y = \beta_0+X\tilde{\beta}_1 + X_{I}\tilde{\beta}_2 + \varepsilon,
\end{equation}
where $X_{I}$ contains all interaction terms, that is, each column of $X_{I}$ is an element-wise product of two columns in $X$, $\tilde{\beta}_1=(\beta_1,\ldots,\beta_p)^T$ is a $p$-dimensional vector of main effects, and $\tilde{\beta}_2=(\beta_{12},\ldots,\beta_{(p-1)p})^T$ is a $\binom{p}{2}$-dimensional vector of interaction effects.
In the simulation, we set $p=10$, $\beta_0=1$, $\tilde{\beta}_1$ and $\tilde{\beta}_2$ vectors of unity, and $\sigma^2=9$.
We investigate the cases when the full sample sizes are $n=5\times10^3, 10^4, 10^5,$ and $10^6$ and the fixed subsample size is $k=1000$.
When the covariates follow a multivariate normal
distribution as in Case 2, the left and middle plots in Figure \ref{interact} present the $\log_{10}$ of $\MSE_{\tilde{\beta}_1}$ and $\MSE_{\tilde{\beta}_2}$ versus $\log_{10}(n)$, and the right plot presents the $\log_{10}$ of $\MSE_{\tilde{\beta}_1}$ if a first-order linear model (without interaction terms) is fitted.
Plots for Cases 1 and 3 are similar and omitted here for brevity.
We observe from the plots that the OSS method consistently outperforms other subsampling methods for parameter estimations. First, the $\MSE_{\tilde{\beta}_1}$ for OSS is always lower than those for other methods, meaning that estimations of main effects derived from the subsample selected by OSS is more accurate than estimations from other subsampling methods. Second, the smaller $\MSE_{\tilde{\beta}_2}$ for OSS demonstrates the ability of its subsample to identify significant interaction effects. Third, the smaller $\MSE_{\tilde{\beta}_1}$ of OSS in the mis-specified first-order model ensures the robustness of its subsample against model mis-specifications and the presence of interactions. These merits of OSS make its subsample a primary choice when potential interactions exist among covariates.

It is known that when all covariates are scaled to $[-1,1]$, an orthogonal array of strength 4 is universally optimal \revb{for the first-order plus interaction model \eqref{c4inter} \citep{dey1999fractional}}. If we define a discrepancy function by
\begin{equation}\label{inter4}
L_4(X_s) = \sum_{1\leq i<j \leq k} \big[p-\|x_i^*\|^2/2-\|x_j^*\|^2/2+\delta(x_i^*,x_j^*)\big]^4,
\end{equation}
its minimization generates an orthogonal array of strength 4, that is, $L_4(X_s)$ is minimized if and only if $X_s$ forms a two-level orthogonal array of strength 4. \reva{This result can be proved similarly as Theorem \ref{c4jeq} using a result from  \cite{xu2003minimum}}. Note that \revb{$L_4(X_s)$} in \eqref{inter4} differs from \revb{$L(X_s)$ in \eqref{c4D}} only on the fourth power at the end of \eqref{inter4}. It seems that, for models with interactions, it makes more sense to use $L_4(X_s)$ in \eqref{inter4} as the discrepancy function in the OSS algorithm instead of $L(X_s)$ in \eqref{c4D}. We have examined the OSS algorithm with $L_4(X_s)$ and find that it tends to provide a subsample of similar performance as with $L(X_s)$. In fact, a full sample rarely contains an orthogonal array subset, so both \revb{$L(X_s)$ and $L_4(X_s)$} are far away from their theoretical minimums. Since the minimization of both discrepancies seeks for extreme data points with dissimilar signs, the two selected subsamples are frequently similar, if not exactly the same, with a limited number of distinct data points.

For linear regression with both interactions and quadratic terms, the OSS approach also performs better than available methods. To further boost the performance of OSS, following the central composite design strategy \citep{box1951experimental}, we can add some center and axial points into the subsample. To keep the subsample size at $k$, we may select $k_0<k$ data points by OSS and select $k-k_0$ data points close to the center or axes.
In our simulation, this modification increases the efficiency of the subsample if the data have a small dimensionality $p$, say, for $p\leq4$, while for moderate and large $p$, the addition of those points usually ends up with a big reduction on the efficiency. {This is because for a larger $p$, the data points selected from OSS cover the center and axial points, as was shown in Figure \ref{proj},} making the addition of those points {redundant}. Therefore, for linear regression with both interactions and quadratic terms, when $p\geq 5$, the OSS approach is still a better choice than available methods for subsample selection.

\subsection{Computing time}
Table \ref{comptime} shows the running times (including the selection and analysis of the subamples, in seconds) modeling with or without interactions under different settings of the full sample size $n$, the number of covariates $p$, and the subsample size $k$.
All computations are carried out on a laptop running macOS 10.15.5 {with 2.2 GHz} Intel Core i7 processor and 16GB memory. \reva{For any setting of $n$, $p$, and $k$ \revb{in Table \ref{comptime}}, OSS is faster than IBOSS \revb{except for one case}, and they are both much faster than modeling on the full data.
Specifically, when $n=10^6$ and $p=500$, or $p=30$ with interaction terms included, modeling on the full sample runs out of the memory while OSS and IBOSS subsamples can both allow us to make statistical inference. For modeling with interactions, OSS is significantly faster than IBOSS because OSS chooses subsamples only based on covariates while IBOSS relies also on the interaction terms.}
It should be noted that IBOSS \citep{wang2019divide} and OSS can both be further accelerated with parallel computing, making the selection and analysis of the subsamples much more efficient.

\begin{table}[t]
  \centering
  \caption{Running times (in seconds) for different combinations of $n$, $p$, and $k$.}
  \begin{tabular}{rrrrr}
    \hline
    \multicolumn{5}{c}{(a) CPU times for different $p$, $n=10^6$, and $k=4p$ for modeling without interactions.}\\
    \hline
    $p$  & \hspace{1cm} UNI & \hspace{2cm} OSS & \hspace{1.8cm} IBOSS & FULL\\
    \hline
    50     & 0.004  &  1.243  & 2.022  &  4.704\\
    100  &  0.009   &  2.086  & 4.270 &  14.310\\
    500  &  0.359   &  13.536 &  24.220 &    ---~~~\\
    \multicolumn{5}{c}{(b) CPU times for different $n$, $p=50$, and $k=1000$ for modeling without interactions.}\\
    \hline
    $n$  & \hspace{1cm} UNI & \hspace{1.5cm} OSS & \hspace{1.2cm} IBOSS & FULL\\
    \hline
    $10^5$  & 0.007   &  0.223  & 0.151 & 0.392\\
    $10^6$  &  0.013  &  1.264   &2.014 & 4.695\\
    $10^7$  &  2.213  &  17.641   &23.660 & 64.468\\
    \multicolumn{5}{c}{(c) CPU times for different $p$, $n=10^6$, and $k=1000$ for modeling with interactions.}\\
    \hline
    $p$ & UNI &OSS & IBOSS & FULL \\
    \hline
    $10$ & 0.008  &  0.448 &  2.263 & 4.871\\
    $20$ & 0.043  &  0.694 &  9.426 &52.421\\
    $30$ & 0.136  &  1.030 &  21.476 & ---~~~\\
    \hline
  \end{tabular}
  \label{comptime}
\end{table}

\subsection{Application to Real Data}\label{sec:realdata}
We evaluate the performance of the OSS approach on two real data examples. In the first example, we examine the accuracy of the OLS estimates of parameters in first-order linear models based on OSS subsamples. In the second example where $p$ is bigger, we consider Lasso regression and examine the prediction accuracy of the models trained on OSS subsamples.

\begin{figure}[t]
  \centering
   \includegraphics[width=.35\textwidth]{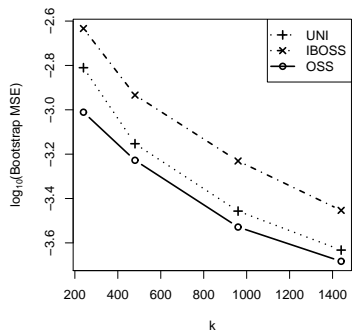}
   \vspace{-3mm}
  \caption{Bootstrap MSEs for estimating slope parameters.}
  \label{wavemse}
\end{figure}

\subsubsection{Wave Energy Converters Data}
This dataset consists of locations and absorbed power of wave energy converters in four real wave scenarios from the southern coast of Australia (Sydney, Adelaide, Perth and Tasmania).
The full dataset has $n=288,000$ data points and contains readings of 32 location variables and 16 absorbed power variables, so the number of covariates in the model is $p=48$. The response variable is the total power output of the farm. Further information on the dataset can be found at ``UCI Machine Learning Repository" \citep{Dua:2019}.  In our analysis, we
work with the log-transformation of the response variable.

Our goal is to consider the accuracy of the OLS estimates of parameters based on subsamples selected via different methods.
We consider the MSE for the vector of slope
parameters by using 100 bootstrap samples, as was done in \cite{wang2018information}.
Each bootstrap sample is a random sample of size $n$ from the full data using uniform sampling with replacement.
For a bootstrap sample, we implement each subsampling method to obtain subsamples and parameter estimates. The bootstrap MSEs are the empirical MSEs and therefore measure the accuracy of parameter estimates.
We consider subsample sizes $k=5p,10p,20p$, and $30p$, and the bootstrap MSEs are plotted in Figure~\ref{wavemse}. The performance of the OSS method dominates other subsampling methods in minimizing the bootstrap MSEs.
The poor performance of IBOSS in this example is probably because that not all variables are important in this dataset. Since IBOSS selects points with the most information in each variable, when some of them are not important, those corresponding points do not contain as much information as expected and will result in a loss of information in the selected subsample. In contrast, the OSS subsample approximates an orthogonal array when projected to any subset of variables (a merit carried over from orthogonal arrays) and therefore performs well when only a subset of variables are important. 

\subsubsection{Blog Feedback Data}
The \revb{data} originate from blog posts. The prediction task associated with the data is the prediction of the number of comments for a blog post in the upcoming 24 hours. To simulate the situation, a basetime (in the past) is set and the blog posts published at most 72 hours before the basetime are selected. The full training dataset has $n=52,397$ data points each of which contains $p=280$ features (variables) of the selected blog posts at the basetime. The features include the characteristics about the source (the blog on which the post appeared) of the post such as the total number of comments for the source before the basetime and the number of comments in the last 24 hours, and the characteristics of the blog post such as the length of the post and the frequencies of some key words.
The basetimes were in the years 2010 and 2011 in the training data. A testing dataset with 7,624 data points is provided where the basetimes were in February and March 2012. This simulates the real-world situation in which training data from the past are available to predict events in the future. It is recommended to use the provided training and testing split to ensure disjoint partitions and thus fair evaluation of the analysing methods.
Further details about the data can be found at \cite{buza2014feedback} and ``UCI Machine Learning Repository" \citep{Dua:2019}.

\begin{figure}
  \centering
  \includegraphics[width=.9\textwidth]{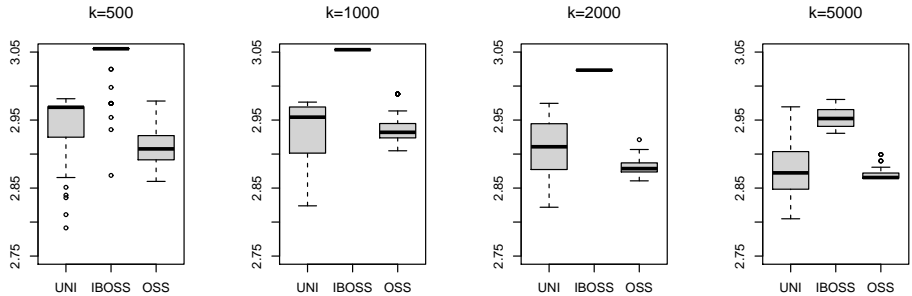}
  \vspace{-5mm}
  \caption{MSPEs of the LASSO regression models for the testing data.}\label{LassoPred}
\end{figure}

Because $p$ is big in this dataset, we use LASSO regression \citep{tibshirani1996regression} to select important features and make predictions.
We consider subsample sizes $k=500,1000,2000$, and 5000 and
select subsamples from the training data with different methods. On each subsample, a LASSO regression model is fitted via the R package ``glmnet" \citep{friedman2010regularization} with the parameter $\lambda$ selected by cross validation. The obtained models are then used to make predictions for the testing data. Figure \ref{LassoPred} plots the mean squared prediction errors (MSPEs) in 100 repetitions of this process. The uniform subsamples select different variables across repetitions and give significantly different predictions, so their performance is not stable at all. The models \reva{trained on} the IBOSS subsamples for $k=500, 1000,$ and 2000 do not select any important variable and thus provide the worst predictions. For $k=5000$, the models based on the IBOSS subsamples select a single variable (the 52th variable, that is, the number of comments for the source in the last 24 hours), so there is some variation on the prediction due to the variation on the coefficient estimation.
The OSS subsamples select more important variables and provide better predictions. For example, for $k=5000$, the OSS subsamples always select the 10th (the median of the number of comments for the source in the last 24 hours), the 21st (the average difference between the number of comments in the last 48 to 24 hours and the number of comments in the last 24 hours for the source), and the 52nd (the number of comments for the source in the last 24 hours) variables. They allow more stable and accurate predictions than other subsamples for any setting of $k$.

\section{Discussion}

In this paper, we propose a new subsampling method, the OSS, for big data linear regression problems. The idea of selecting subsamples resembling orthogonal arrays is justified by the universal optimality of orthogonal arrays. A key appeal of orthogonal arrays is their combinatorial orthogonality, which provides a balance among data points and covariates, maximizing the overall information provided by data points.
The OSS method exploits this orthogonality and efficiently selects subsample points that best approximate an orthogonal array. Simulations confirm the improved performance of the subsample selected by the OSS method over other available subsampling methods, and the practical applicability of the method is illustrated using \reva{two} real-world applications. An efficient C++ implementation of OSS is available as an R package.

A greedy modification of Algorithm \ref{alg} is possible and may bring us better subsamples. To do this, after a subsample of size $k$ is obtained from the algorithm, randomly choose a subsample point $x_i^*$ and replace it with a point $x$ in the full sample if $l(x|X_{s,(-i)})<l(x_i^*|X_{s,(-i)})$, where $l$ is the discrepancy in \eqref{relaloss} and $X_{s,(-i)}$ is the current subsample only without $x_i^*$. The algorithm can stop until no such $x$ can be found in the full sample. Theoretical properties on convergence may also be examined for this greedy algorithm. While such a greedy version may enhance the performance of the subsample on parameter estimations, the computing time will also grow dramatically. Considering that the primary usage of subsampling methods is to accelerate the analysis of big data, cumbersome subsampling approaches are \revb{neither} applicable nor valuable.
Another possible modification of Algorithm \ref{alg} is to obtain multiple subsamples by randomly selecting some initial points from the full sample and then choose the best subsample with the largest $D-$ or $A$-efficiencies. The increase in computing time would depend on the number of initial points. Although this modification may bring about a better subsample for inference purposes, it shares the same computing issue as the greedy version.
{While the current Algorithm \ref{alg} does not guarantee the minimization of the discrepancy function in \eqref{c4D} or the convergence of the selected subsample to an orthogonal array, our extensive numerical results are strong evidence that the algorithm tends to efficiently produce a dramatically improved subsample relative to those found by other methods.}

Though Algorithm \ref{alg} is focused on subsampling for linear regression problems, it can be readily modified to select subsamples for other study objectives because it is a general and efficient algorithm for approximating combinatorial optimality. Given the combinatorial nature of all subsampling tasks, we only need to change the discrepancy function used in the algorithm to accommodate different subsampling objectives. For example, there are several discrepancy functions for various modeling techniques or model free tasks \citep{fang2006design} and  they all can be directly used in the algorithm to select subsamples that target different goals of data analysis. The development of new discrepancy functions for new subsampling objectives is also meaningful and requires further study.

%
%

 \section*{Acknowledgements}
 The authors would like to thank an associate editor and reviewers for their helpful comments and suggestions.

%

\begin{supplement}
The supplementary materials \citep{Supplement2021} include the proof of Theorem \ref{c4jeq}
and a figure showing the performance of the OSS method on the estimation of intercept in a first-order linear model.
\end{supplement}


\bibliographystyle{imsart-nameyear} 
\bibliography{ref}       


\end{document}